\documentclass[copyright,creativecommons]{eptcs}
\usepackage[T1]{fontenc}
\usepackage[utf8]{inputenc}
\usepackage{float}
\usepackage{amsmath}
\usepackage{graphicx}

\makeatletter

\pdfpageheight\paperheight
\pdfpagewidth\paperwidth

\providecommand{\tabularnewline}{\\}
\floatstyle{ruled}
\newfloat{algorithm}{tbp}{loa}
\providecommand{\algorithmname}{Algorithm}
\floatname{algorithm}{\protect\algorithmname}

\newenvironment{lyxcode}
{\par\begin{list}{}{
\setlength{\rightmargin}{\leftmargin}
\setlength{\listparindent}{0pt}
\raggedright
\setlength{\itemsep}{0pt}
\setlength{\parsep}{0pt}
\normalfont\ttfamily}%
 \item[]}
{\end{list}}

\makeatother

\begin{document}
\title{The Glasgow Parallel Reduction Machine: Programming Shared-memory Many-core Systems using Parallel Task Composition}

\author{Ashkan Tousimojarad, Wim Vanderbauwhede\institute{School of Computing Science\\University of Glasgow\\Glasgow, UK}\email{a.tousimojarad.1@research.gla.ac.uk, wim@dcs.gla.ac.uk}}

\def\titlerunning{GPRM: Programming Shared-memory Many-core Systems}

\def\authorrunning{A. Tousimojarad, W. Vanderbauwhede}

\maketitle

\begin{abstract}
We present the Glasgow Parallel Reduction Machine (GPRM), a novel,
flexible framework for parallel task-composition based many-core programming.
We allow the programmer to structure programs into task code, written
as C++ classes, and communication code, written in a restricted subset
of C++ with functional semantics and parallel evaluation. In this
paper we discuss the GPRM, the virtual machine framework that enables
the parallel task composition approach. We focus the discussion on
GPIR, the functional language used as the intermediate representation
of the bytecode running on the GPRM. Using examples in this language
we show the flexibility and power of our task composition framework.
We demonstrate the potential using an implementation of a merge sort
algorithm on a 64-core Tilera processor, as well as on a conventional
Intel quad-core processor and an AMD 48-core processor system. We
also compare our framework with OpenMP tasks in a parallel pointer
chasing algorithm running on the Tilera processor. Our results show
that the GPRM programs outperform the corresponding OpenMP codes on
all test platforms, and can greatly facilitate writing of parallel
programs, in particular non-data parallel algorithms such as reductions.
\end{abstract}

\section{Introduction}

As processor clock speeds have stagnated, processor manufacturers
have moved to many-core devices in an attempt to perpetuate Moore's
law. Processors with tens of cores are already commercially available
and soon will be the norm, even in laptops. However, most programming
languages were originally intended for single-core processors and
consequently most software is written for single-core processors.
Efficient utilisation of many-core platforms is a great challenge.
POSIX threads enable parallel programming but they are difficult to
use and put the burden on the programmer, even when using compiler
directives such as OpenMP; there are number of languages where parallelism
can be expressed natively without the need for explicit thread creation
\cite{weiland2007chapel,thies2002streamit,Armstrong2007PES,pointon2001design},
but compared to mainstream languages such as C++ and Java, none of
them have found widespread adoption. Even if a multicore programming
language would find wide adoption, it would in the short term obviously
be impossible to rewrite the vast amount of single-core legacy code
libraries, nor would it be productive. For many applications, especially
computationally-intensive ones, sequential algorithms are extremely
efficient. We therefore propose an approach to parallel programming
based on parallel composition of (sequential) tasks. Our task composition
approach can be integrated into existing code and has the advantage
of providing a parallel task composition mechanism embedded in an
existing language.

In order to express irregular parallelism, the concept of \textit{tasking}
is introduced in OpenMP 3.0 \cite{ayguade2009design}. Although with
the OpenMP tasks, one can express more parallelism, or write parallel
codes more easily, it sometimes results in a situation in which lots
of fine-grained tasks are created and the performance degrades dramatically
\cite{podobas2010comparison}. We will show such a situation later
in section \ref{sec:Example2:-Parallel-pointer}. 

Intel Threading Building Blocks (TBB) is a famous approach for expressing
task-based parallelism \cite{reinders2010intel}. Intel TBB is a C++
runtime library that contains data structures and algorithms to simplify
parallel programming. It abstracts the low-level threading details
required to utilise the multi-core systems, similar to the approach
that we are using. However, there are still some important issues
that put burden on the programmer, such as dealing with the mutual
exclusion. 

Cilk++ \cite{leiserson2010cilk++}, which is a variation of Cilk that
supports C++, uses some keywords to express task parallelism. In the
Cilk++ the scheduling of tasks are predefined, while our approach
allows for different task scheduling strategies to be defined statically
or dynamically at run-time.

The SMP superscalar (SMPSs) project from the Barcelona Supercomputing
Center \cite{superscalar2008user,perez2008dependency} also allows
programmers to write sequential applications and the framework is
able to exploit the existing concurrency and to use the different
processing cores by means of an automatic parallelisation at run time.
The SMPSs runtime builds a data dependency graph where each node represents
an instance of an annotated function and edges between nodes denote
data dependencies. The SMPSs program code must be annotated using
special preprocessor directives. 

Our approach is to provide a language with default parallel evaluation
implemented using a restricted subset of C++ which is familiar and
easy to use for the end users.

\section{A Task Composition Framework for Many-core Platforms}

To facilitate reuse of existing single-core code libraries we propose
a \emph{task-based} approach to many-core programming, in particular
for computationally intensive tasks. In our parlance, a \emph{task
node} consists of a \emph{task kernel} and a \emph{task manager}.
A \emph{task kernel} is typically a complex, self-contained unit offering
a specific functionality. Such a kernel is said to provide one or
more \emph{services} to the system. A \emph{task kernel} on its own
is not aware of the rest of the system. The \emph{task manager} provides
the task composition interface to the kernel. 

To implement this paradigm, we have created the Glasgow Parallel Reduction
Machine (GPRM), a lightweight, distributed reduction engine for shared-memory
and NUMA-style many-core and multiprocessor systems. The GPRM executes
a strict functional machine language with concurrent evaluation of
function arguments. As the name suggests, the GPRM is similar in spirit,
if architecturally very different, to reduction machines such as Alice
\cite{harrison1987parallel}. The concepts behind reduction machines
are very well explained in \cite{vree1989design}. Essentially, the
GRPM is a coarse-grained reduction machine as we only reduce the task
composition graph. The GPRM uses $\beta$-reduction and performs string
reduction at bytecode level.

Conceptually, the GPRM consists of a set of \emph{tiles} connected
over a network. Each \emph{tile} consists of a \emph{task node} and
a FIFO queue for incoming packets. Every \emph{tile} runs in its own
thread and blocks on the FIFO. The system is event driven, with two
possible types of events: arrival of a packet and events generated
by the kernel. The latter is either creation of a packet or modification
of the local state.

The reduction engine (i.e. the task manager) evaluates the GPIR bytecode
via parallel dispatch of packets requesting computations to other
\emph{tiles}. 

\begin{figure}

\begin{centering}
\includegraphics[width=0.6\columnwidth]{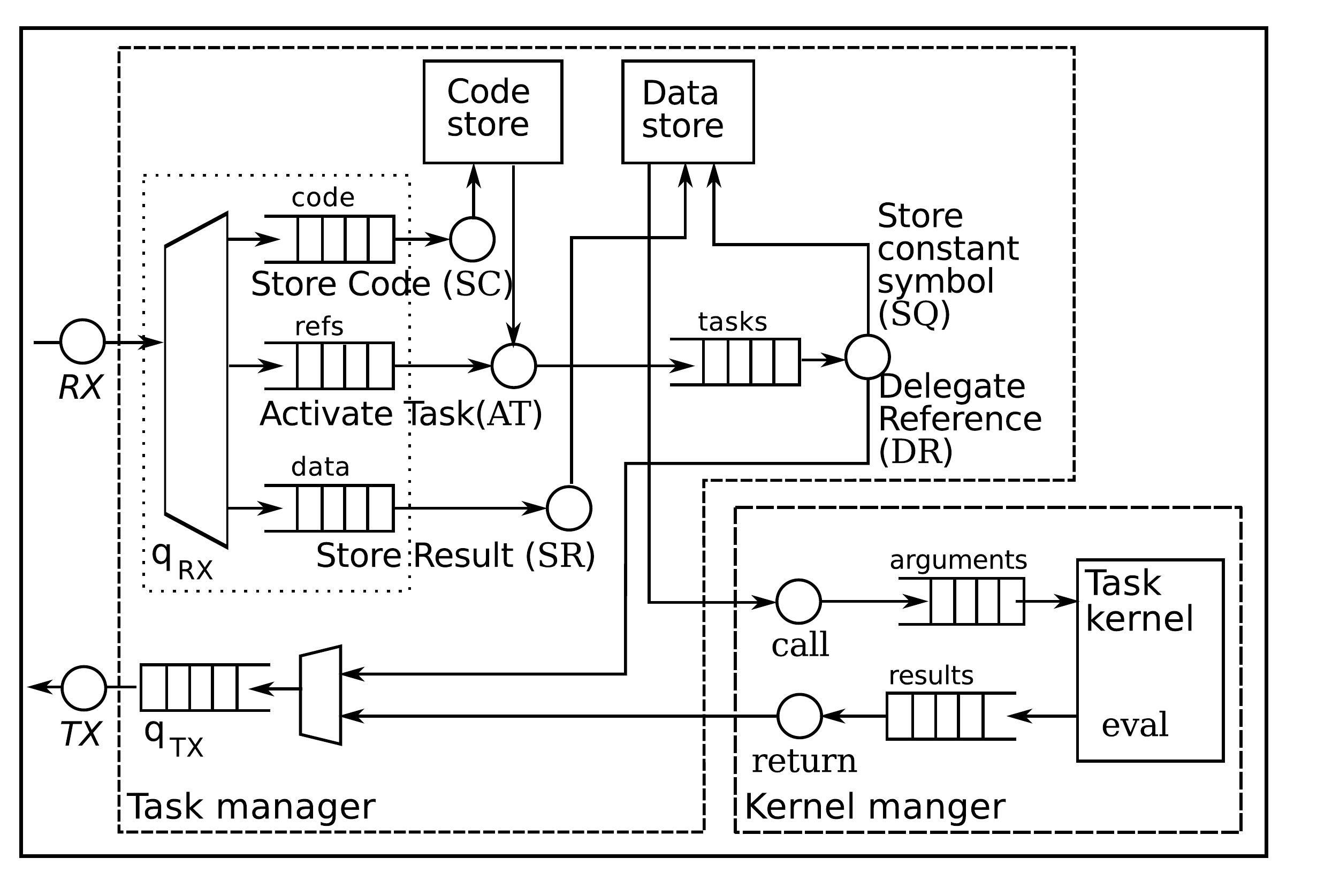}
\par\end{centering}

\caption{\label{fig:GPRM-Task-Manager}GPRM Task Manager}

\end{figure}

The Glasgow Parallel Reduction Machine (GPRM) is a Virtual Machine
in the sense that it evaluates bytecode; it could equally be considered
a runtime library as every instance of the GPRM is compiled based
on the source program, and linked with the original source. The GPRM
evaluates the bytecode in parallel, as follows: 
\begin{itemize}
\item On startup, the GPRM creates a thread pool. 

\begin{itemize}
\item The threads exchange packets via FIFO queues. 
\item Each thread runs a \emph{tile}.
\end{itemize}
\item Computations are triggered by the arrival of a \emph{reference} packet,
a packet which contains a reference to a \emph{subtask}, i.e. a piece
of bytecode representing an S-expression. 

\begin{itemize}
\item Each argument in this S-expression is either a reference or a value. 
\item References are sent out to other tiles for computation, values are
stored. 
\item The leaf subtasks of the computational tree have either no arguments
or constant values as arguments, so no references need to be sent.
\end{itemize}
\item Once all arguments have been evaluated, the reduction engine passes
the evaluated arguments of the S-expression to the task kernel which
performs the actual computation. 
\item The result of the computation is returned to the caller, i.e. the
sender of the reference packet. 
\end{itemize}
Rather than the conventional stack used by other virtual machines,
we use a random-access \emph{subtask list} to store the state of each
parsed subtask. The addresses of the entries in this list (the \emph{subtask
records}) are managed using a stack. The reduction engine keeps track
of which arguments have been evaluated using the \emph{subtask records}:
arrival of a reference packet results in creation of a new subtask
record in the subtask list. The address is taken from the \emph{subtask
stack}. The process is shown in pseudo-code in Algorithm \ref{alg:Subtask-allocation-and}(a).
When the computation by the kernel is finished, the process is essentially
to dispatch the result and clean up, as shown in Algorithm \ref{alg:Subtask-allocation-and}(b).

\begin{algorithm}
\begin{lyxcode}
\medskip{}

\#~(a)~On~receipt~of~subtask~reference

{\small addr~=~subt\_stack.pop()}{\small \par}

{\small subt\_list{[}addr{]}=~subt\_rec.new()}{\small \par}

{\small \#~caller~id~is~taken~from~the~request~packet}{\small \par}

{\small subt\_list{[}addr{]}.caller~=~caller\_id}{\small \par}

{\small subt\_list{[}addr{]}.operation~=~bytecode.shift()}{\small \par}

{\small for~arg~in~bytecode:}{\small \par}

{\small{}~~if~arg.kind==reference}{\small \par}

{\small{}~~~status~=~requested~}{\small \par}

{\small{}~~~val~=~nil}{\small \par}

{\small{}~~~dispatch\_ref\_packet(subt\_addr,~arg)}{\small \par}

{\small{}~~else}{\small \par}

{\small{}~~~status~=~present}{\small \par}

{\small{}~~~val~=~arg~~~}{\small \par}

{\small{}~~end}{\small \par}

{\small{}~~subt\_list{[}addr{]}.args{[}arg{]}=(val,status);}{\small \par}

{\small end}{\small \par}

~

\#~(b)~After~computation~by~kernel

{\small dispatch\_result\_packet(caller\_id,~result)}{\small \par}

{\small subtask\_stack.push(addr)}{\small \par}

\medskip{}

\end{lyxcode}
\caption{\label{alg:Subtask-allocation-and}(a) Subtask allocation and parsing;
(b) Result dispatch and clean-up}
\end{algorithm}

\begin{lyxcode}

\end{lyxcode}
The combined operation of all reduction engines in all threads results
in the parallel reduction of the entire program. Because the communication
between the tasks is expressed using a pure functional language, the
Church-Rosser theorem \cite{church1936some} guarantees that the parallel
evaluation is correct, and we can easily express complex communication
patterns between the tasks. The programmer does not need to deal with
creating and managing threads: parallel execution is the default and
the GPRM manages the thread pool.

\section{Programming Model}

Essentially, the GPRM programming model is one of computational tasks
that communicate using a function call mechanism. The function call
tree is evaluated in parallel. This mechanism can support various
types of parallelism such as data parallelism, reduction, and pipeline
parallelism.

\subsection{Kernel Tasks and Wrapper Architecture}

The computational kernel tasks are written as C++ classes. This means
that the end user simply creates classes in the \emph{GPRM::Kernel::}namespace.
The only requirement on the class is that it does not instantiate
another class in the \emph{GPRM::Kernel::}namespace. Execution of
code in these classes follows C++ semantics, including the possibility
to create threads etc.

The compiler wraps every user class in a pure functional interface
which provides the actual \emph{task} abstraction, essentially by
generating a switch/case statement to select methods corresponding
to byteword values.

\subsection{Communication Code}

The communication between the tasks is expressed in a restricted subset
of C++11 with parallel evaluation. This communication code is compiled
into the Glasgow Parallel Intermediate Representation language (GPIR),
a small, S-expression based functional language.

In practice, the communication code is a function in the GPRM:: namespace,
called from a (sequential) C++ program. The restrictions are mainly
determined by the requirement that the communication code must be
pure and functional, so they imply single assignment and no dynamic
memory allocation. However, because the wrapper functions around the
user classes can actually contain arbitrary C++ code, there are very
few other restrictions. Examples are given below.

\section{Glasgow Parallel Intermediate Representation Language}

The Glasgow Parallel Intermediate Representation (GPIR) is based on
the untyped lambda calculus \cite{pierce2002types}, with extensions
similar to Scheme's \cite{Sussman75scheme:an}: numbers, conditionals,
lists. GPIR is more regular than Scheme in that every expression must
be either a constant, a lambda variable or an operation-operands sequence,
i.e. any list must always start with an operation. Furthermore, the
GPIR has one additional syntactic construct, the quote '. Quoting
an expression defers evaluation from the Reduction Engine to the task
kernel. This \emph{deferred evaluation} is the basic mechanism used
in the GPRM to implement control features. The GPIR syntax is given
by:

~

\emph{<s-expr>} S-expression~of~the~form

\emph{'('~operation~<$expr_{1}$>~...~<$expr_{m}$>~')'}

\emph{operation} A literal (in practice representing the instance
of a class and the method used, e.g. $t_{1}.m_{2})$)

\emph{<}$expr_{i}$\emph{>} Either an S-expression, or a literal (e.g.
\emph{x} or \emph{42}) that is not an operation. Can be quoted, i.e.
preceded by a quote, e.g.: $'(t_{1}.m_{1}\;'42)$

~

\subsection{Compilation of GPIR into Bytecode\label{Compilation-of-GPIR}}

The GPIR compiler converts nested S-expressions into a map of flat
S-expressions by substituting a reference for every non-literal expression,
and then converts the flat S-expressions into bytecode, essentially
by assigning a 64-bit number to every reference and literal. The keys
in the map are the memory addresses where the bytecodes are stored.
For example,
\begin{lyxcode}
{\small{}~}{\small \par}

{\small (t1.m2~}{\small \par}

{\small{}~~(t2.m3~'42)~}{\small \par}

{\small{}~~(t3.m4)}{\small \par}

{\small )}{\small \par}

{\small{}~}{\small \par}
\end{lyxcode}
is converted into a map:
\begin{align*}
 & r_{1}\Rightarrow(t_{1}.m_{2}\, r_{2}\, r_{3})\\
 & r_{2}\Rightarrow(t_{2}.m_{3}\,'42)\\
 & r_{3}\Rightarrow(t_{3}.m_{4})
\end{align*}

The GPIR compiler creates a packet which contains the reference to
the root of the computation (i.e. $r_{1}$ in the example) and sends
it to the corresponding tile for evaluation.

\subsection{Minimal GPIR Subset}

For the purpose of this paper, we consider the GPIR subset consisting
of following types of expressions:

\subsubsection{Lambda expression}

~

\[
(\lambda'x_{1}'x_{2}...'x_{n}'<s\text{–}expr>)
\]

All arguments are quoted to defer evaluation, because evaluation of
a lambda expression by the task manager would be meaningless as the
arguments $x_{i}$ are not constants or references, so their evaluation
is not defined. The compiled bytecode contains a list of flat S-expressions
representing the original nested expression. For example, the lamdba
expression 

\[
(\lambda\,'x\,'(*(-\, x\,'1)\,(+\, x\,'1)))
\]

is compiled into
\begin{align*}
 & r_{1}\Rightarrow(\lambda\,'x\,'r_{2}\,'r_{3}\,'r_{4})\\
 & r_{2}\Rightarrow(*\, r_{3}\, r_{4})\\
 & r_{3}\Rightarrow(-\, x\,'1)\\
 & r_{4}\Rightarrow(+\, x\,'1)
\end{align*}

\subsubsection{Beta reduction expression}

~

\[
(\beta<\lambda-expr><expr_{1}><expr_{2}>...<expr_{n}>)
\]

The beta reduction, i.e. substitution of the lambda variables by their
corresponding argument values, is performed at the level of the bytecode.
After beta reduction, the bytecode will be evaluated by the Reduction
Engine. There are two points worth noting: 
\begin{itemize}
\item First, the arguments are evaluated in parallel. However, achieving
sequential evaluation is very simple: it suffices to convert

\[
(\lambda'x_{1}'x_{2}...'x_{n}'<s\text{–}expr>)
\]

into

\[
(\lambda'x_{1}'(\lambda'x_{2}'(\lambda\;...'(\lambda'x_{n}'<s\text{–}expr>))...))
\]

\item Second, if the expressions are not quoted, they will be evaluated,
i.e. they will return values. However, as GPRM tasks are C++ objects,
the methods will return either numerical values (e.g. int or float)
or pointers. The wrapper methods return these wrapped in a bytecode
container, so that the beta-substitution results in valid GPIR bytecode.
However, if the expressions are quoted, then the quoted byteword is
obviously valid bytecode; the quote is removed during the substitution,
so that the references will be evaluated.
\end{itemize}

\subsubsection{Conditional expression}

\emph{
\[
({\it if}\,{\it <cond\text{–}expr>}\;{\it '<if\text{–}true\text{–}expr>}\;'<{\it if\text{–}false\text{–}expr}>)
\]
}In the \emph{if} expression, if the \emph{if-true} and \emph{if-false}
expression are quoted, only the expression corresponding to the value
of the condition is evaluated.

\subsubsection{List expressions}

~

The list expressions are defined in the usual functional style, based
on the empty list and the cons operation.

\subsubsection{Expression labels}

Any expression in the GPIR can be explicitly labeled and called by
its label:\emph{
\begin{align*}
({\it label}\, L & <{\it expr_{1}}>)\\
(op_{2}\; L)
\end{align*}
}is equivalent to\emph{
\[
({\it op_{2}}\,<{\it expr_{1}}>)
\]
}

\subsubsection{Additional expressions}

For reasons of efficiency, in practice there is a larger set of GPIR
expressions, for example to support sequencing of operations without
the need for chained function calls. However, all of these can theoretically
be expressed in terms of the above minimal set. For example:

~

\emph{(return <expr>)= (if '1 '<expr> '0)}

~

\emph{(begin <expr$_{1}$><$expr_{2}$>...<$expr_{n}$>) = }

\emph{($\beta\;$($\lambda'x_{1}...'x_{n}\;$'(return$\; x_{n}$))<$expr_{1}$><$expr_{2}$>...<$expr_{n}$>)}

~

\emph{(let (assign 'x <x-expr>) '<in-expr>) =}

\emph{($\beta$ ($\lambda$ 'x '<in-expr>) <x-expr>)}

~

\subsection{Compiling GPRM C++ Code to GPIR}

Code for the GPRM is C++ code. We use the name GPC for the language
of the sections of the code that are executed on the GPRM. The GPC
compiler separates the task code from the communication code based
on the namespace. In this paper, the GPC language is not our primary
concern. As a simple example, the following code 

~
\begin{lyxcode}
{\small GPRM::Kernel::Task1~t1;}{\small \par}

{\small GPRM::Kernel::Task2~t2;}{\small \par}

~

{\small int~GPRM::compute(int~v0)~\{}{\small \par}

{\small int~v1~=~t1.m1(v0);}{\small \par}

{\small int~v2~=~t2.m1(v1);~}{\small \par}

{\small int~v3~=~t2.m2(v1);~}{\small \par}

{\small return~t1.m2(v2,v3);}{\small \par}

{\small \}}{\small \par}

~
\end{lyxcode}
will be compiled into GPIR as

~

\textit{($\beta$ }

\textit{~~~~($\lambda$ 'v1}

\textit{~~~~~~'(t1.m2 }

\textit{~~~~~~(t2.m1 v1)}

\textit{~~~~~~(t2.m2 v1)}

\textit{~~~~))}

\textit{~~(t1.m1 (ctrl.arg '0)))}

~

where \textit{(ctrl.arg '0)} is a method of the \emph{ctrl} kernel
to handle the arguments. 

The compilation is conceptually straightforward: because of the restrictions
imposed on the communication language, its abstract syntax is that
of a functional language, and hence it can be compiled easily into
GPIR. The object instance declarations are mapped to tiles based on
the dependencies in the call tree: tasks that depend on one another
can't run in parallel and hence can be mapped onto the same tile.

\subsection{Build Process}

The object instances, together with system information -- in particular
the maximum number of hardware threads in the system -- are used to
dimension the thread pool. Furthermore, the compiler analyses all
the classes and methods used in the task description code and maps
them to numeric constants. These constants are used in the wrapper
function to match the operation from the GPIR code with the actual
method call to be executed. This task-specific generated code is combined
with the generic GPRM code and the source code for the task classes
and compiled into a library. The original program is adapted by adding
an instance declaration for the GPRM and replacing the call to the
task code (GPRM::compute in the above example) by a call to the GPRM's
run method, with the name of the compiled bytecode file as an argument.
The program is then compiled and linked with the GPRM library.

\subsection{Scheduling of Work on Threads }

Dynamic scheduling of work on threads is performed using a control
kernel: whereas a task \emph{(t1.m1 ...) }will be executed on a compile-time
assigned thread, using the \emph{ctrl.run} service, a task can be
scheduled on a run-time computed thread: \emph{(ctrl.run '(t1.m1 ...)
(thread-id-expression)).} The \emph{thread-id-expression }can be a
compile time constant, a run-time computed value or a value returned
at run time by a dynamic scheduler.

\subsubsection{Deferring Evaluation}

The mechanism used to perform the scheduling is based on the GPRM's
ability to defer evaluation to the kernels by quoting. The bytecode
for the above example is:
\begin{align*}
r_{1} & \Rightarrow({\it ctrl.run\,'r_{2}\, r_{3})}\\
r_{2} & \Rightarrow(s_{1}.m_{1}\,\ldots)\\
r_{3} & \Rightarrow({\it thread\text{–}id\text{–}expression})
\end{align*}

The evaluation of $r_{1}$ results in two argument values: the quoted
reference $'r_{2}$ and the number \emph{thread-id. }

\subsubsection{Restarting Evaluation}

The discussion in Section\ref{Compilation-of-GPIR} glossed over the
actual structure of the reference byteword: apart from containing
the address of the corresponding bytecode, it also contains the identifier
of the tile on which this code should be run, in other words the reference
is a tuple of two integers: 
\[
r=({\it code\text{–}addr,}tile\text{–}id)
\]

The \emph{ctrl.run} kernel substitutes the compile-time computed \emph{tile-id}
for $r_{2}$ by the run-time computed value, i.e. \emph{thread-id}.
It then removes the quote and restarts the evaluation. As a result,
a reference packet is dispatched to the run-time computed tile.

\section{Example: Parallel Merge Sort}

To illustrate the programming model, we consider an implementation
of the Merge Sort algorithm which is a good example of parallel reduction.
We will discuss different aspects of this algorithm in detail. In
the next section, we will illustrate the power of the GPRM for task
management using a pointer chasing algorithm.

\subsection{GPRM Implementation}

We use two tasks, \emph{leaf} and \emph{stem}, implemented as methods
of a MergeSort class. The GPC task composition code uses a recursive
tree: 
\begin{lyxcode}
{\small GPRM::Kernel::MergeSort~ms;}{\small \par}

~

{\small void~ms\_rec(int~n,int~nmax,~int{*}~a)~\{}{\small \par}

{\small{}~~if~(n>=nmax)~\{}{\small \par}

{\small{}~~~~ms.leaf(n,a);~}{\small \par}

{\small{}~~\}~else~\{}{\small \par}

{\small{}~~~~ms.stem(}{\small \par}

{\small{}~~~~~~ms\_rec(2{*}n,nmax),}{\small \par}

{\small{}~~~~~~ms\_rec(2{*}n+1,nmax),}{\small \par}

{\small{}~~~~~~a);}{\small \par}

{\small{}~~\}}{\small \par}

\}

{\small int{*}~GPRM::merge\_sort~(int{*}~a)~\{}{\small \par}

{\small{}~~ms\_rec(1,NUM\_THREADS,a);~}{\small \par}

{\small{}~~return~a;}{\small \par}

{\small \}}{\small \par}

~
\end{lyxcode}
In GPIR, this becomes:

~
\begin{lyxcode}
\textrm{\textit{($\beta$~}}

\textrm{\textit{	($\lambda$~'f~'n~'nmax~'a~($\beta$~f~n~nmax~a))~}}

\textrm{\textit{	($\lambda$~'n~'nmax~'a~~~~~~~~~~~~~}}

\textrm{\textit{~~~~~~~~~~~~(if~(>=~n~nmax)}}

\textrm{\textit{~~~~~~~~~~~~~~~'(ctrl.run~}}

\textrm{\textit{~~~~~~~~~~~~~~~~~'(ms.leaf~n~a)~~}}

\textrm{\textit{~~~~~~~~~~~~~~~~~~n)}}

\textrm{\textit{~~~~~~~~~~~~~~~'(ctrl.run~}}

\textrm{\textit{~~~~~~~~~~~~~~~~~'(ms.stem~}}

\textrm{\textit{~~~~~~~~~~~~~~~~~~~~($\beta$~f~({*}~'2~n)~nmax)}}

\textrm{\textit{~~~~~~~~~~~~~~~~~~~~($\beta$~f~(+~'1~({*}~'2~n))~nmax)~)~}}

\textrm{\textit{~~~~~~~~~~~~~~~~~~~~a)}}

\textrm{\textit{~~~~~~~~~~~~~~~~~n)}}~~~~~~~~~

\textrm{\textit{~~~~~~~~~~~~~~~)}}

\textrm{\textit{~~~~~~~~~~)}}

\textrm{\textit{'1~~NUM\_THREADS~(ctrl.reg~'0))}}

~
\end{lyxcode}
For simplicity, we omitted the GPIR code for argument handling. Note
the explicit run-time thread allocation, which allocates the computations
\emph{ms.leaf} and \emph{ms.stem} to the GPRM nodes with address \emph{n}. 

\begin{figure}[h]
\begin{centering}
{\footnotesize }%
\begin{tabular}{cc}
\includegraphics[clip,width=0.45\textwidth]{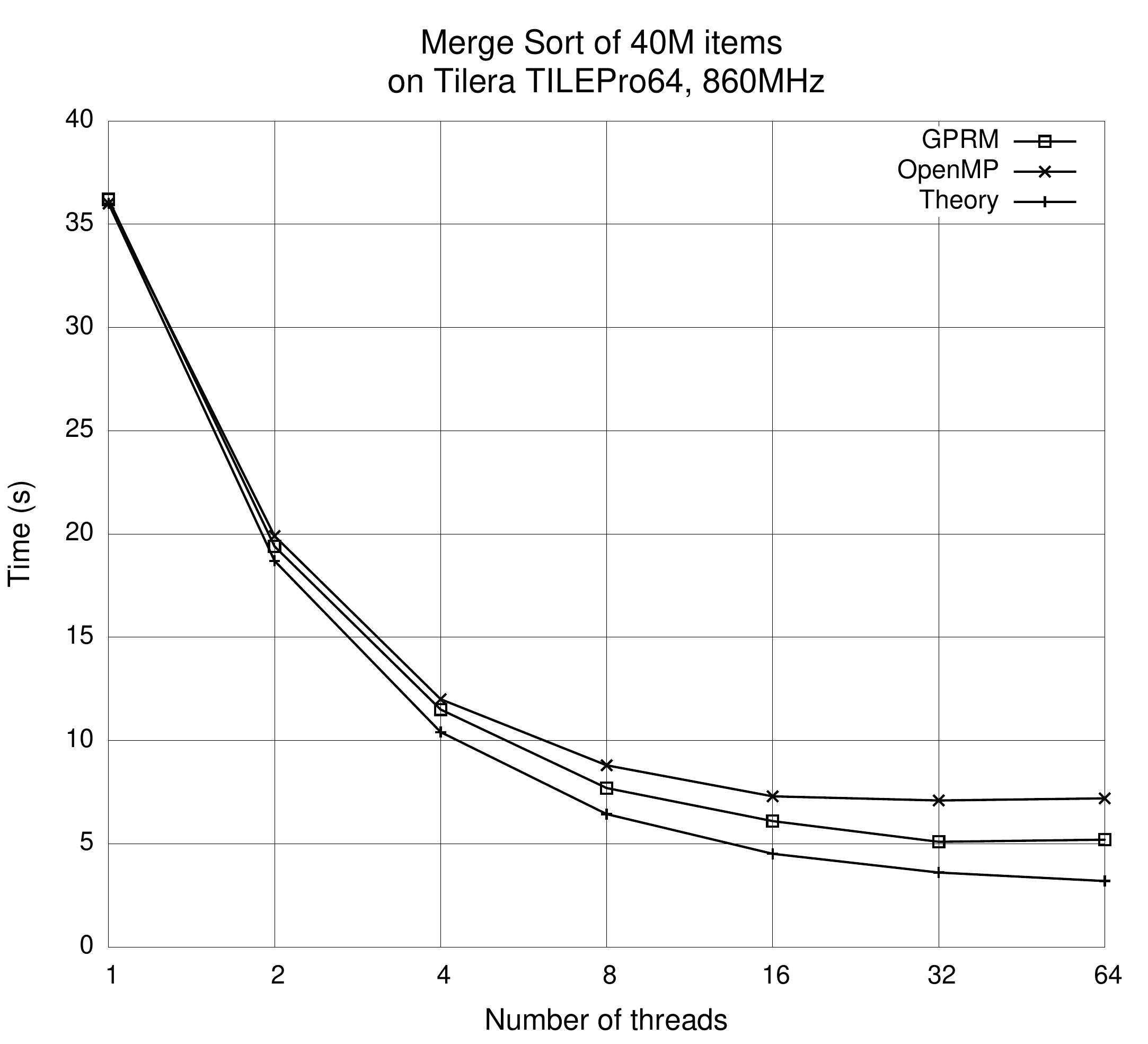} & \includegraphics[clip,width=0.45\textwidth]{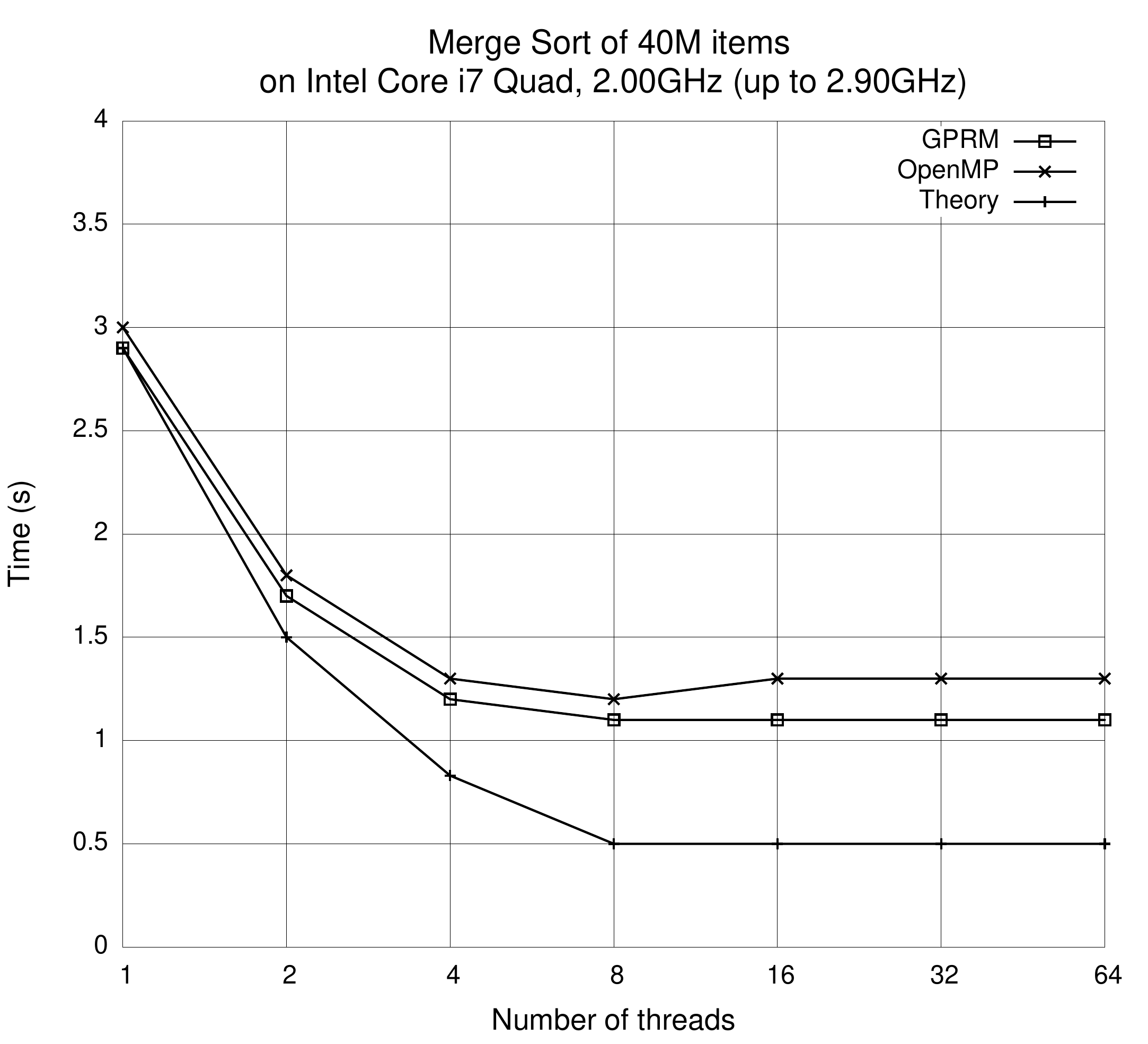}\tabularnewline
{\footnotesize (a)} & {\footnotesize (b)}\tabularnewline
\end{tabular}
\par\end{centering}{\footnotesize \par}

\noindent \begin{centering}
\includegraphics[clip,width=0.5\textwidth]{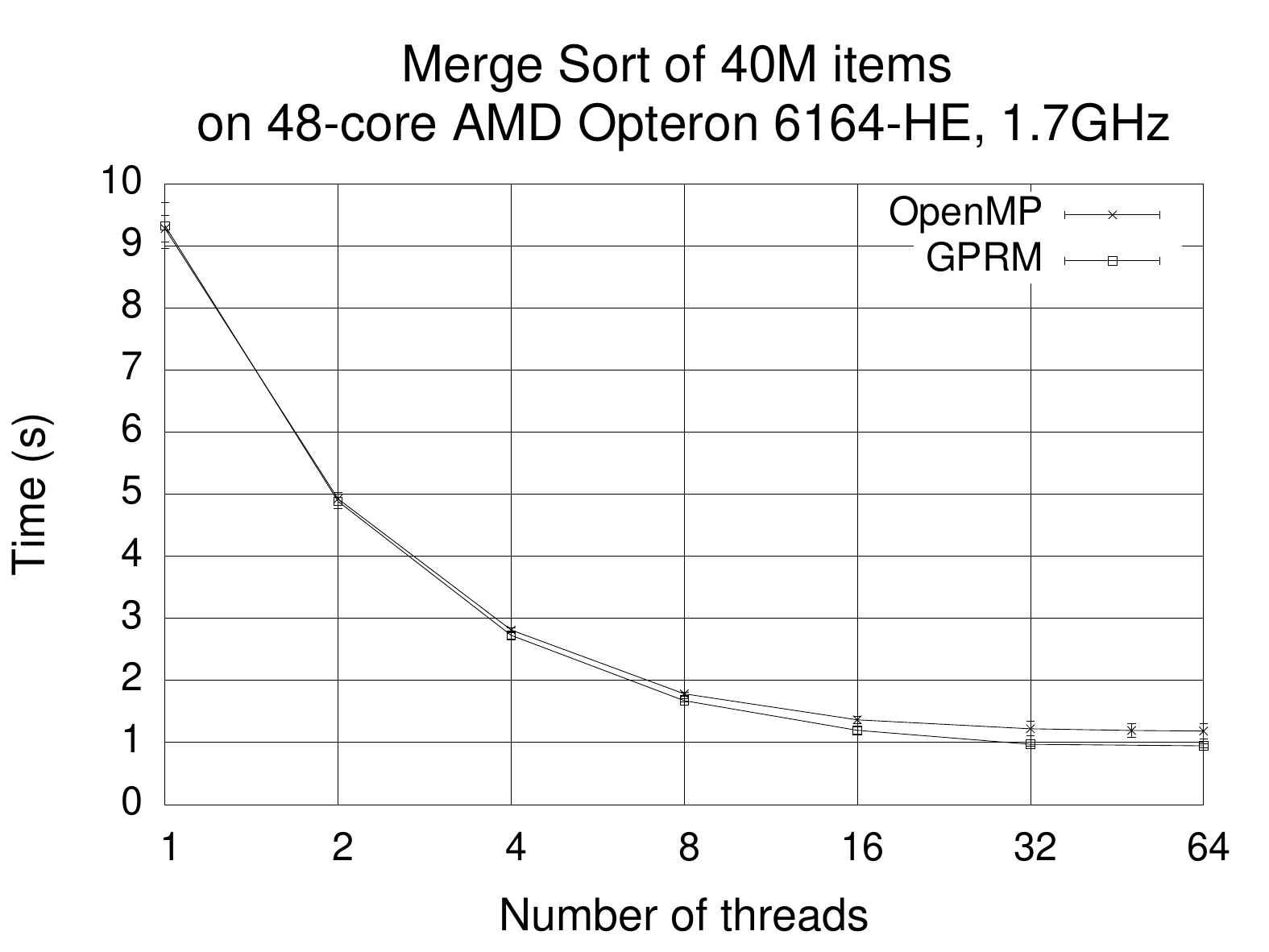}
\par\end{centering}

\begin{centering}
{\footnotesize (c)}
\par\end{centering}{\footnotesize \par}

\caption{\label{fig:Performance-of-merge-1-1}Performance of merge sort in
GPRM and OpenMP on (a) Tilera TILEPro64, (b) Intel Core i7 Quad CPU
and (c) AMD Opteron 6164-HE}
\end{figure}

\subsection{Evaluation and Discussion}

\subsubsection{Parallel Execution Time }

Figure \ref{fig:Performance-of-merge-1-1} shows the performance of
the GPRM compared to a high-performance OpenMP implementation \cite{radenskishared}
for sorting an array of 40M 32-bit integers. We used the Tilera TILEPro64
64-core processor \footnote{http://www.tilera.com/products/processors/TILEPRO64},
as well as on two more conventional platforms: a quad-core Intel Core
i7-2630QM running at 2 GHz, with Turbo Boost up to 2.9 GHz, and a
4-socket, 12-core AMD Opteron 6164 HE, running at 1.7 GHz. Note that
the Tilera processor runs at only 860MHz. Differences in CPU speed
and memory specifications account for the difference in performance
at low numbers of threads. The theoretical model shown in Figure \ref{fig:Performance-of-merge-1-1}
is based on the fact that the execution time for a single-threaded
Merge Sort algorithm is $k.n.log(n)$, in which $k$ is the time required
for each operation and $n$ is the number of elements. Therefore,
the execution time on e.g. 2 cores will be $k\frac{n}{2}\log\frac{n}{2}+kn$.
On the TILEPro64, the maximum number of available cores is 63 (one
is required for the PCIe link between the card and host), which would
result in a critical path and thus some irregularity at the end of
the plot, compared to the theoretical model. Also, with the hyper-threading
technology, the number of available cores on the Intel Core i7 platform
is 8 which restricts the performance gain at higher numbers of threads. 

It is clear from these results that the GPRM is very competitive on
all platforms, indeed the GPRM slightly outperforms the OpenMP version
for larger numbers of threads. Furthermore, the example illustrates
how easy it is to create complex parallel programs using this approach.
The corresponding OpenMP code (see Appendix) uses parallel OpenMP
sections to assign recursive calls to threads. It is noticeably longer
and more complex than the GPC code.

\subsubsection{GPRM Overhead}

\begin{figure}[H]
\begin{centering}
\includegraphics[width=0.6\columnwidth]{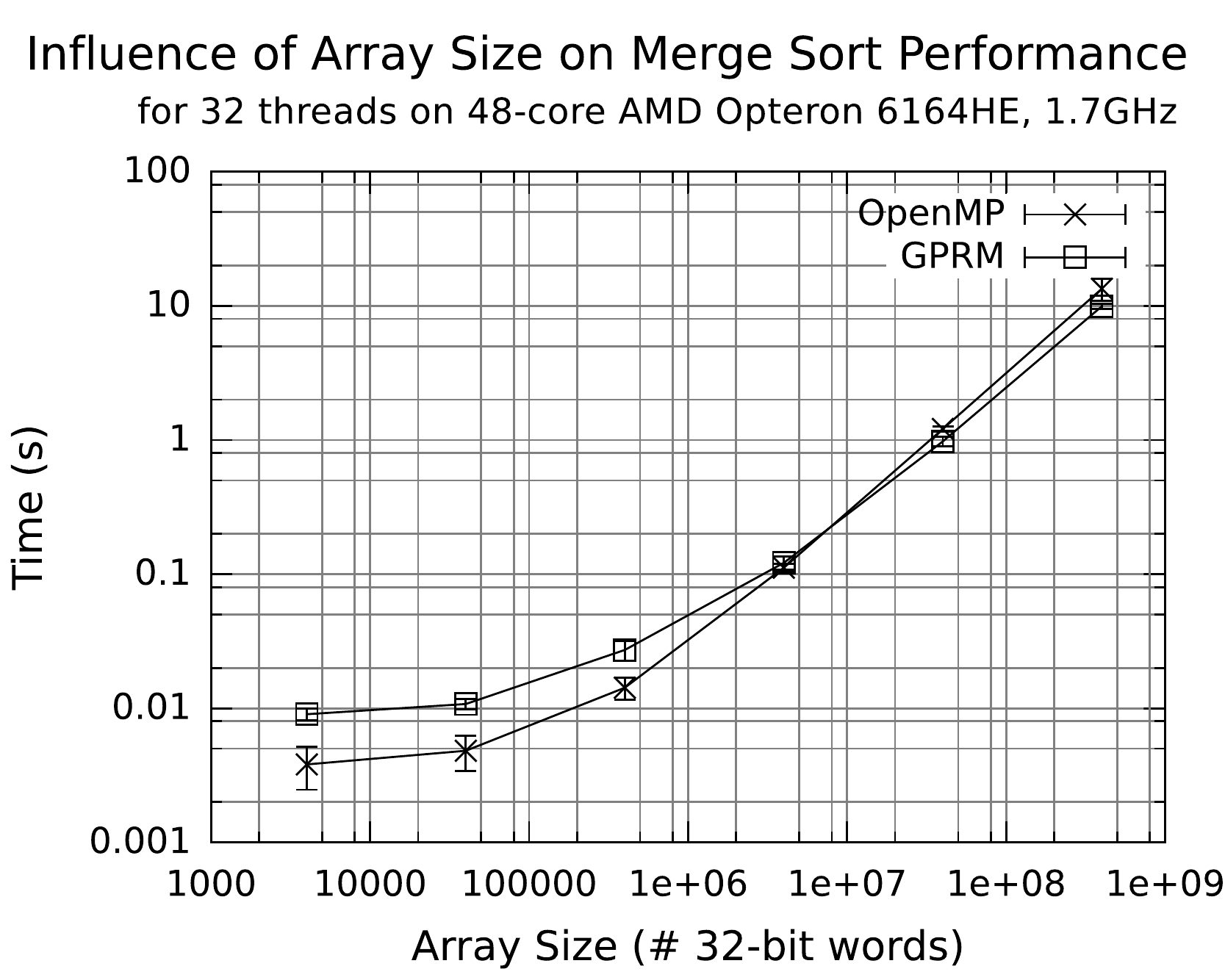}
\par\end{centering}

\caption{\label{fig:Influence-of-Array}Influence of Array Size on Performance}
\end{figure}

As the GPRM is a virtual machine interpreting bytecode, it introduces
some overhead. To explore this we varied the array size from 4K words
to 400M words. In this experiment, we used 32 threads on the 48-core
AMD system. As can be seen from Figure \ref{fig:Influence-of-Array},
for very small arrays (4K) the GPRM code is 2.4$\times$ slower than
the OpenMP code. For 4M, the performance is the same and for 400M,
the GPRM is 1.4$\times$ faster. We are confident that we can reduce
the overhead of the GPRM as up to now no special care was taken to
optimise the performance of the reduction engine.

\subsubsection{Impact of Caching}

\begin{figure}[H]
\begin{centering}
\includegraphics[width=0.6\columnwidth]{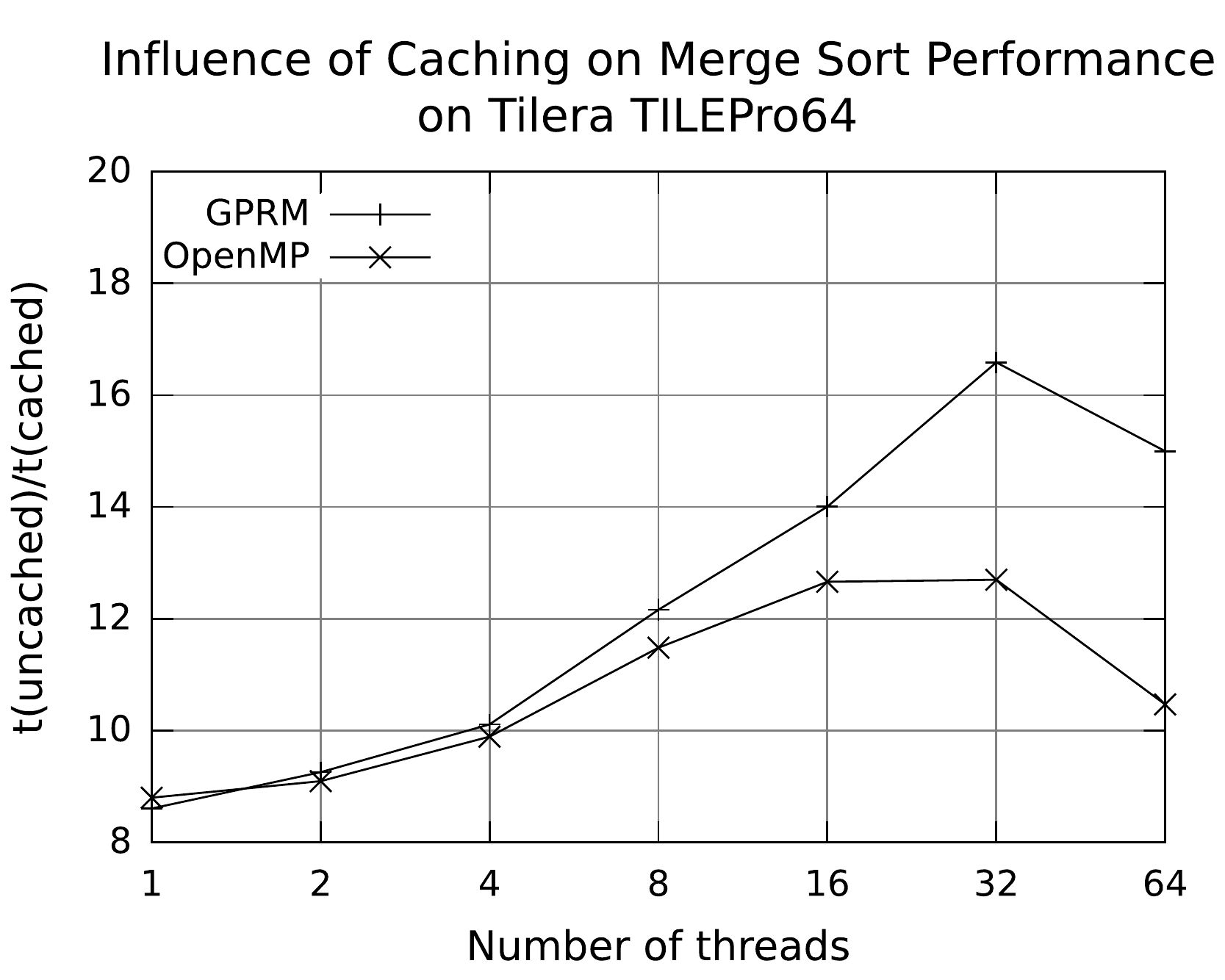}
\par\end{centering}

\caption{\label{fig:Speedup-cache}Influence of Caching on performance on Tilera
TILEPro64}
\end{figure}

An important point is the effect of the memory architecture on the
performance: typically, the off-chip memory (DRAM) has limited parallelism,
so any parallelism provided by the cores could be undone by the memory
bottleneck. To evaluate this, we used the TILEPro64 with caching disabled.
The Tilera platforms provide fine-grained control over caches and
memory, making them ideal for this type of experiments \cite{wentzlaff2007chip}.
Figure \ref{fig:Speedup-cache} shows the ratio of the execution time
for uncached to cached execution: for the GPRM the performance without
caching is 8$\times$ worse for a single thread, and 17$\times$ for
large numbers of threads. We also see that the GPRM makes better use
of the caches than OpenMP, which results in better performance for
larger numbers of threads.

\section{Example2: Parallel pointer chasing\label{sec:Example2:-Parallel-pointer}}

As another example, we compare the performance of our framework with
different OpenMP implementations of a parallel pointer chasing algorithm.
In \cite{ayguade2009design}, pointer chasing algorithm is stated
as one of the motivations behind tasking implementation in OpenMP.
We use two different approaches that are used in that paper to show
how they scale compared to the GPRM implementation. The first OpenMP
approach uses the \texttt{single nowait} construct inside a \texttt{parallel}
region. Each thread needs to traverse the whole list and determine
in each step whether another thread processed the current element
or not. In order to do some work at each element, we have used the
Ackermann function \cite{sundblad1971ackermann}. In the second OpenMP
approach, a single thread traverses the list and creates one task
for each element.

\begin{figure}[H]
\begin{centering}
\includegraphics[width=0.6\columnwidth]{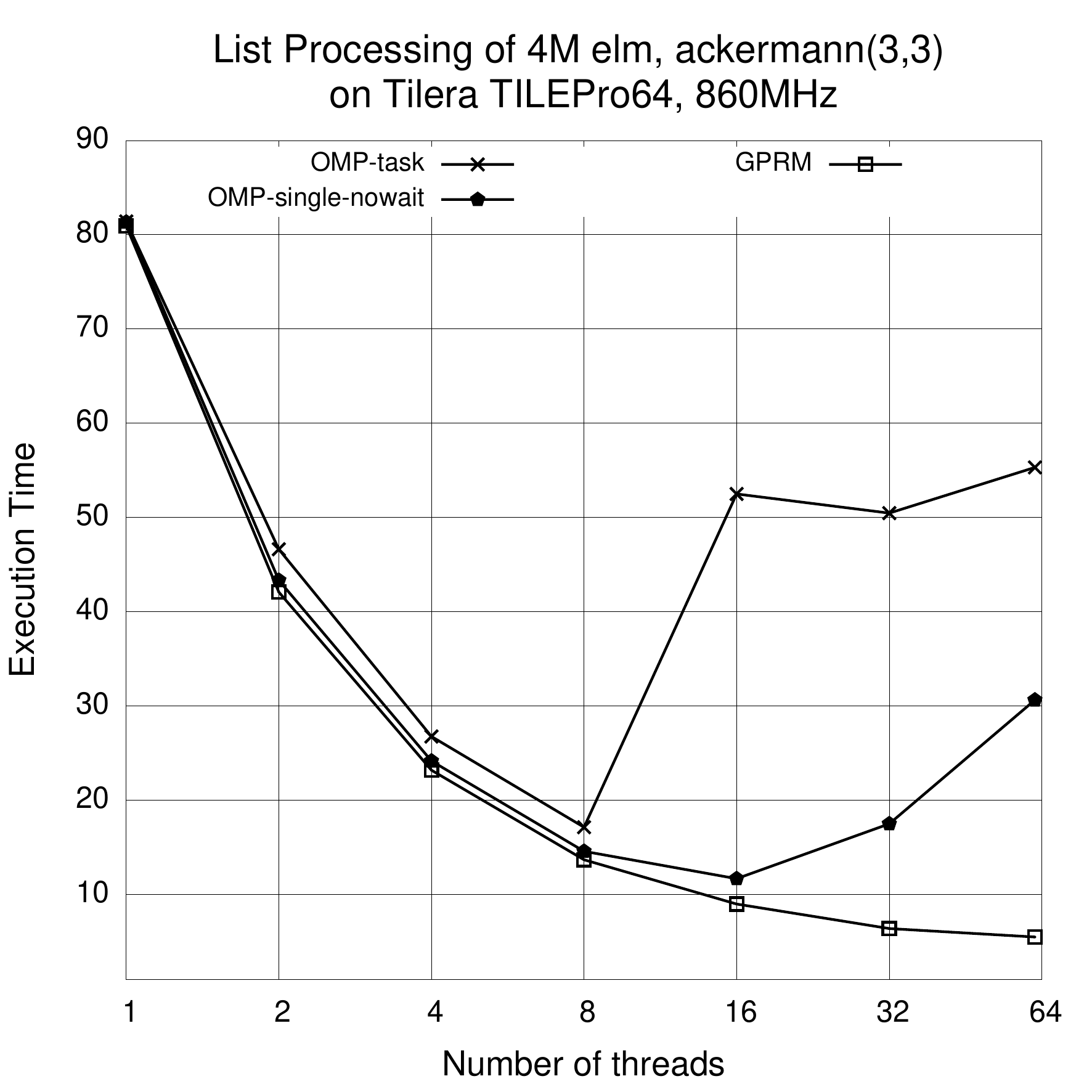}
\par\end{centering}

\caption{\label{fig:Ackermann}Performance of list processing in GPRM and two
OpenMP implementation on the Tilera TILEPro64}
\end{figure}

Figure \ref{fig:Ackermann} shows that GPRM scales very well when
the number of threads becomes larger. It is also evident that the
OpenMP tasking approach performs poorly when so many tasks are created.
Although in the GPRM implementation, every thread traverses the whole
list, it does not need to check whether other threads have processed
the element or not. The reason is that the whole work is divided between
threads in such a way that there is no contention between them. For
example the thread k is responsible to process the element k, NTH+k,
2NTH+k, and so on.

\section{Conclusion and Future Work}

We have presented a new approach towards many-core programming, based
on a functional task composition language with parallel reduction,
implemented as the Glasgow Parallel Reduction Machine. We have shown
that our approach slightly outperforms OpenMP for the merge sort algorithm
on large arrays. For the pointer chasing algorithm, we have demonstrated
that out approach works so much better than the corresponding OpenMP
implementations. It does not have the high cost of the \texttt{single}
construct, or the overhead of having so many fine-grained tasks. We
are currently working on a data-parallel version of the GPRM, for
use on GPUs and, eventually, FPGAs. In this way we aim to create a
unified programming framework for heterogeneous many-core systems.

\paragraph*{Acknowledgement}

We thank Prof. Hameed of the University of Aizu, Japan, for kindly
granting us the use of the AMD system for our experiments. Ashkan
Tousimojarad would like to thank SICSA (Scottish Informatics and Computer
Science Alliance) for supporting his research.

\bibliographystyle{eptcs}
\bibliography{paper_EPTCS2013_GPRM}
\newpage{}

\section*{Appendix 1. OpenMP Code for Merge Sort }
\begin{lyxcode}

{\small }
\begin{algorithm}[H]
\begin{lyxcode}
\medskip{}

{\small \#include~<omp.h>}{\small \par}

{\small{}~}{\small \par}

{\small int~mergesort\_serial(int{*}~input,~}{\small \par}

{\small{}~~int{*}~scratch,~int~size);}{\small \par}

{\small{}~}{\small \par}

{\small void~merge(~int{*}~input1,~}{\small \par}

{\small{}~~int~size1,~int{*}~input2,~}{\small \par}

{\small{}~~int~size2,~int{*}~scratch);}{\small \par}

{\small{}~}{\small \par}

{\small int~mergesort\_parallel\_omp~(int{*}~input,~}{\small \par}

{\small{}~~int{*}~scratch,~int~size,~int~threads)~\{}{\small \par}

{\small{}~~~~~if~(threads~==~1)~\{}{\small \par}

{\small{}~~~~	~int~r~=~mergesort\_serial(input,~}{\small \par}

{\small{}~~~~~~~~~~~scratch,~size);}{\small \par}

{\small{}~~~~	~return~r;}{\small \par}

{\small{}~~~~~\}}{\small \par}

{\small{}~~~~~else~\{}{\small \par}

{\small \#pragma~omp~parallel~sections}{\small \par}

{\small{}~~~~~~\{}{\small \par}

{\small \#pragma~omp~section}{\small \par}

{\small{}~~~~	~~\{}{\small \par}

{\small{}~~~~	~~~mergesort\_parallel\_omp(input,~}{\small \par}

{\small{}~~~~~~~~~~~~~scratch,~size/2,~threads/2);}{\small \par}

{\small{}~~~~	~~\}}{\small \par}

{\small \#pragma~omp~section}{\small \par}

{\small{}~~~~	~~\{}{\small \par}

{\small{}~~~~	~~~mergesort\_parallel\_omp(input+size/2,~}{\small \par}

{\small{}~~~~~~~~~~~~~scratch+size/2,~size-size/2,~}{\small \par}

{\small{}~~~~~~~~~~~~~threads-threads/2);}{\small \par}

{\small{}~~~~	~~\}}{\small \par}

{\small{}~~~~~~\}}{\small \par}

{\small{}~~~~~~merge(input,~size/2,~}{\small \par}

{\small{}~~~~~~~input+size/2,~size-size/2,~scratch);}{\small \par}

{\small{}~~~~\}}{\small \par}

{\small{}~~return~0;~}{\small \par}

{\small \}}{\small \par}

{\small{}~}{\small \par}

{\small int~main()~\{}{\small \par}

{\small{}~~~~omp\_set\_nested(1);}{\small \par}

{\small{}~~~~omp\_set\_num\_threads(2);}{\small \par}

{\small{}~~~~int{*}~array1~=~new~int{[}ARRAY\_SZ{]};}{\small \par}

{\small{}~~~~int{*}~scratch1~=~new~int{[}ARRAY\_SZ{]};}{\small \par}

{\small{}~~~~mergesort\_parallel\_omp(array1,~}{\small \par}

{\small{}~~~~~~scratch1,~ARRAY\_SZ,~NUM\_THREADS);}{\small \par}

{\small{}~~return~0;}{\small \par}

{\small \}}{\small \par}

\medskip{}

\end{lyxcode}
{\small \caption{Parallel Merge Sort using OpenMP sections \cite{radenskishared}}
}
\end{algorithm}
{\small \par}

\end{lyxcode}

\end{document}